\documentclass[reprint, amsmath,amssymb,aps]{revtex4-1}
\usepackage{graphicx}
\usepackage{amsmath}
\usepackage{amsfonts}
\usepackage{amssymb}
\usepackage{color}
\usepackage{dcolumn}

\newcommand{\beq}{\begin{equation}}
\newcommand{\eneq}{\end{equation}}
\newcommand{\bea}{\begin{eqnarray}}
\newcommand{\enea}{\end{eqnarray}}

\begin{document}

\title{Resolving the effects of frequency dependent damping and quantum phase diffusion in YBa$_2$Cu$_3$O$_{7-x}$ Josephson junctions}

\author{D. Stornaiuolo$^1$}
\email[present address:]{ Department of Condensed Matter Physics, University of Geneva 24 Quai E.-Ansermet,
CH-1211 Geneva 4, Switzerland; daniela.stornaiuolo@unige.ch}
\author{G. Rotoli$^2$}
\author{D. Massarotti$^{3,1}$}
\author{F. Carillo$^4$}
\author{L. Longobardi$^{2,1}$}
\author{F. Beltram$^4$}
\author{F. Tafuri$^{2,1}$}

\affiliation{$^{1}$CNR-SPIN Napoli, Complesso Universitario di Monte Sant'Angelo, Napoli, Italy,\\$^{2}$Dip. di Ingegneria Industriale e dell'Informazione Seconda Universit\`{a} di Napoli, Aversa (CE), Italy,\\$^{3}$Universit\`{a} degli Studi di Napoli Federico II, Italy, \\$^{4}$NEST, Scuola Normale Superiore, I-56126 Pisa,Italy}

\date{\today}

\begin{abstract}
We report on the study of the phase dynamics of high critical temperature superconductor Josephson
junctions. We realized YBa$_2$Cu$_3$O$_{7-x}$ (YBCO) grain boundary (GB) biepitaxial junctions in the submicron scale, using low loss substrates, and analyzed their dissipation by comparing the transport measurements with Monte Carlo simulations. The behavior of the junctions can be fitted using a model based on two quality factors, which results in a frequency dependent damping. Moreover, our devices can be designed to have Josephson energy of the order of the Coulomb energy. In this unusual energy range, phase delocalization strongly influences the device's dynamics, promoting the transition to a quantum phase diffusion regime. We study the signatures of such a transition by combining the outcomes of Monte Carlo simulations with the analysis of the device's parameters, the critical current and the temperature behavior of the low voltage resistance $R_0$.    
\end{abstract}



\maketitle

\section{Introduction}

A correct understanding of the phase dynamics of a Josephson circuit relies on the possibility to distinguish the contributions to dissipation coming from the junction itself from those due to the external circuit. This is especially relevant in the moderately damped regime for junctions with low critical current. High temperature superconductor (HTS) Josephson junctions (JJ) often fall in this category. Their phase dynamics is made particularly rich by the HTS unconventional superconductivity \cite{kirtley, vanharlingen, tafuri_RPP}. The high value of the critical temperature (T$_c$ $\approx$90 K) and of the superconducting gap  ($\Delta$ $\approx$ 20 meV) impose a unique energy scale to HTS JJs. Some effects generally observed in HTS junctions, as for example the values of the $I_cR_N$ parameter (with $I_c$ and $R_N$ the critical current and normal state resistance respectively) on average one order of magnitude lower than the expected value of 2$\Delta$, may signify the relevance of other energy scales in these devices \cite{ tafuri_RPP, gross_97, reviewGB}. One possibility is the Thouless energy associated to single nanoscale channels in a filamentary approach to transport across the GB \cite{lucignano_PRL10}.

Despite this complexity, recent experiments demonstrate that macroscopic quantum phenomena can be observed also in HTS JJs,\cite{bauch_PRL, bauch_science, inomata} revealing coherence beyond expectations.  Ultrasmall HTS junctions were also used to realize single electron transistors with unprecedented energy resolution,\cite{gustafsson} and proposed for the fabrication of ultra-sensitive superconducting quantum interference devices to use in the detection of small spin systems.\cite{koelle_condmat, koelle_sust} These studies confirm the interest in nanoscale HTS devices and the need for a systematic and reliable study of their phase dynamics.

A detailed analysis of phase dynamics in moderately damped low temperature superconductor (LTS) JJs was performed by Kautz and Martinis in the early 90s\cite{kautz_martinis}. Here it emerges the need of a frequency dependent damping to fully account the phenomenology of the junctions, with clear indications of distinct behaviors at low and high frequency respectively. These arguments offer the possibility to disentangle the quality factor of the junction from the one of the external circuit. More recently, moderately damped JJs based on both LTS and HTS and operating in the phase diffusion regime, were investigated through the analysis of the switching current distribution (SCD) histograms\cite{phase_diffusion, Yu_PRL11, Longobardi_PRL12}. All these devices are, however, characterized by values of the Josephson energy $E_J=\hbar I_0/2e$  (where $I_0$ is the critical current in absence of thermal fluctuations) much larger than those of the charging energy $E_c=e^2/2C$ (where $C$ is the junction capacitance). Devices characterized by $E_J \approx E_c$, on the other hand, were first studied by Iansiti et al.\cite{iansiti} using Sn based junctions with nominal area of $\sim$0.1$\mu$m$^2$ and $I_c$ in the range 1-10nA. It was shown that this energy scale favours the access to a quantum phase diffusion regime, which is quite unexplored and whose nature is still unsettled.\cite{Yu_PRL11, Longobardi_PRL12, iansiti, tinkham}  
\\
In this work we study the phase dynamics of sub-micron HTS JJs in the moderately damped regime, using the tools developed for LTS JJs. We have realized YBCO junctions with lateral size down to 600nm on (La$_{0.3}$Sr$_{0.7}$)(Al$_{0.65}$Ta$_{0.35}$)O$_3$ (LSAT)  substrates. The reduction of the junctions' size allows one to minimize the influence of the GB microstructure on the transport properties of the devices,\cite{koelle_sust, submicron, stornaiuolo_SUST} while the use of LSAT substrate reduces the parasitic capacitance present in the more common SrTiO$_3$ (STO) based junctions.\cite{stornaiuolo_JAP} Using Monte Carlo simulations, we extract the frequency dependent damping of these devices and show that, for a particular range of parameters, the quantum phase diffusion regime can be attained.

\section{Experimental}
The junctions studied in the present work were realized following the design reported in Ref.[\onlinecite{stornaiuolo_SUST, nuovo, ceo2}]. A CeO$_2$ thin film is deposited using RF magnetron sputtering on a (110) oriented LSAT substrate and patterned using photolithography and ion-beam etching (IBE). A 200nm YBCO film is then deposited using inverted cylindrical magnetron sputtering, obtaining (001) growth on the CeO$_2$ seed layer and (103) growth on the LSAT substrate, and subsequently covered with a protective gold layer (100nm thick). The definition of the sub-micron bridges is carried out using an electron beam lithography technique adapted to HTS requirements \cite{carillo_PRB10}. The electron beam pattern is transferred to a 80-nm-thick Ti layer which serves as a hard mask. The YBCO not covered by the Ti mask is removed using IBE, keeping the sample at low temperature (-140$^\circ$C) in order to minimize oxygen loss. After this, the Ti mask is removed by chemical etch in a highly diluted (1:20) HF solution. Finally, the protective gold layer is removed using a last step of low-energy IBE. In the panels (a) and (c) of Fig. 1 scanning electron microscope images of 600nm wide devices (before the gold removal) are shown. The high quality of the YBCO film can be inferred from the systematic presence of elongated grains with typical size of 1$\mu$m in the (103) part and by the absence of impurities and outgrowths in the (001) part.\cite{TEM_99} 
\\

\begin{figure}[tbp]
\centering
        \includegraphics[width=8.5 cm, height=12.5 cm]{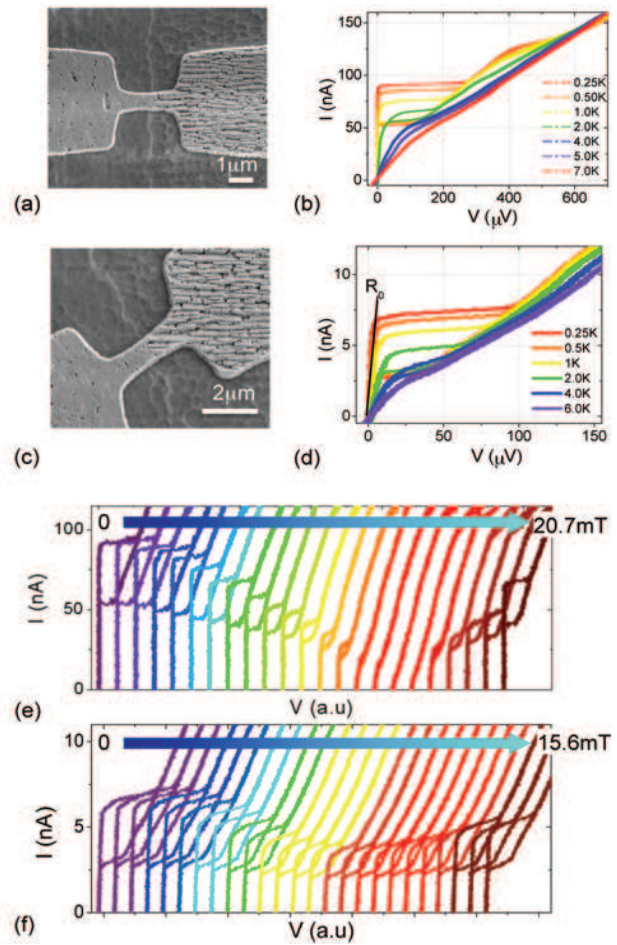} 
    \caption{(Color online) SEM images of devices 1W (a) and 6W (c) and plots of the relative $I-V$ characteristics (panels (b) and (d) respectively) measured at various temperatures. The width is 600nm for both devices. Panels (e) and (f) show the $I-V$s measured at T=0.25K as a function of $H$ (applied in the junction's barrier plane) for device 1W and 6W respectively. The curves were shifted horizontally for clarity. $H$ is ramped from 0 to 20.7mT in steps of 0.9mT in panel (e) and from 0 to 15.6mT in steps of 0.6mT in panel (f).}
\end{figure}
 
The devices were measured down to 0.25K using a four contacts technique. The measurement environment was magnetically shielded and the lines were filtered using RC filters and two stages of copper powder filters.\cite{luigi_filters} Current vs. voltage ($I-V$) characteristics of two typical devices, 1W and 6W, are shown in panels (b) and (d) of Fig. 1 respectively. The $I-V$s are modulated by the magnetic field $H$ (panels (e) and (f)), leading to a Fraunhofer-like $I_c(H)$ pattern for junction 1W.\cite{bp, tafuri_RPP}. Taking into account focusing effects,\cite{rosenthal} the $I_c(H)$ pattern periodicity in field points to an effective width of $\approx$500nm for device 1W (Figure 1e) and of $\approx$600nm for device 6W (Figure 1f). These values are very close to the nominal dimensions of the devices. The critical current density $J_c$ is 65 A/cm$^2$ for device 1W and 5 A/cm$^2$ for device 6W.
The low $J_c$ values of these devices are a consequence of oxygen depletion, occurring especially in the GB region.\cite{oxygenLSAT} This is a quite general feature of HTS JJs\cite{tafuri_RPP} and is expected to be of particular relevance when decreasing the size of the junction, as in this case. We have found that the devices realized using LSAT as a substrate are characterized by higher values of the normal state resistance and are more affected by aging when compared with the ones fabricated on STO substrates. These micro-structural factors could in this case mask the influence of the d-wave order parameter in determining the magnitude of the $J_c$ as a function of the junction misorientation.\cite{nuovo} Grains elongated in the current direction in the device 1W (Fig. 1(a)), for instance, might be less exposed to oxygen desorption compared to grains leaning against the walls of the channel in device 6W (Fig. 1(c)), explaining the different values of $I_c$ measured for these two devices. 

The reduced values of $J_c$, on the other hand, offers the possibility to have access to JJ dynamical regimes which have been poorly explored. The Josephson energy $E_{J}$ is $\approx$ 270 $\mu$eV (corresponding to 3K) for device 1W and 70$\mu$eV  (corresponding to 0.8K) for device 6W. These energies were calculated using the $I_0$ values obtained through comparison to numerical results, as described in Section IV. They are one or two orders of magnitudes smaller than those measured for junctions where macroscopic quantum behavior has been demonstrated\cite{bauch_science}, and five orders of magnitude smaller than those observed in most HTS Josephson devices.\cite{tafuri_RPP, reviewGB} More importantly, for device 6W, $E_{J}$ is comparable with the charging energy $E_{c}$, as will be described in Section IV, placing this device in an uncommon and interesting energy range.
\\

\begin{figure}[t]
\centering
        \includegraphics[width=5.5 cm, height=4.0 cm]{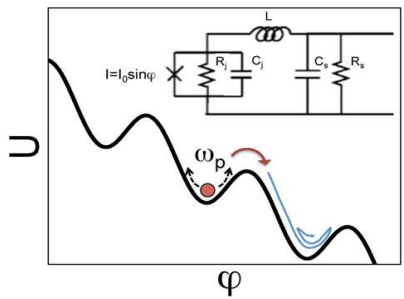} 
    \caption{(Color online) Tilted "washboard" potential of a Josephson junction. The red arrow indicates the effect of thermal activation and the blue one the recapturing of the fictitious phase particle in overdamped junctions. In the inset, the circuit considered in the frequency dependent damping model is shown.}
\end{figure}
 
The $I-V$ curves shown in Fig. 1 are highly hysteretic, with a difference between the critical ($I_c$) and the retrapping ($I_r$)  current up to 70$\%$ at the lowest temperature (panel d). The presence and the nature of hysteresis in the $I-V$s of HTS junctions have been a matter of debate.\cite{tafuri_RPP} It is indeed difficult in these devices to disentangle the intrinsic capacitive effects in the GB barrier from extrinsic ones, deriving from the external circuit, also due to the high dielectric constant (above 10000 at low temperatures\cite{dielconstSTO}) of the STO substrates on which the junctions are commonly fabricated.\cite{tafuri_RPP, reviewGB, bauch_science, rotoli_PRB07} In the present work, we have used LSAT as a substrate, with a temperature independent dielectric constant $\epsilon_{r}$ of 23 \cite{dielconst_LSAT}. As a consequence, the influence of the external circuit is greatly reduced.\cite{stornaiuolo_SUST} 
\\
Remarkably, this neat hysteresis coexist with a slope at low voltage. The low voltage slope is an hallmark of phase diffusion effects \cite{kautz_martinis} and is visible in Fig. 1(b) (device 1W) for temperatures greater than 2K, and in Fig. 1(d) (device 6W) in the whole temperature range, down to the 0.25K. The two phenomena, hysteresis and phase diffusion, can separately be understood in the framework of the washboard potential model for Josephson junctions.\cite{bp} On the other hand, their coexistence in the same $I-V$ is unusual \cite{kautz_martinis, ono_IEEE87, iansiti} and requires a finer analysis of the devices properties and dynamics, which we will address in the following section.

\section{The "tilted washboard" potential model for the study of JJ's phase dynamics}
The behavior of a Josephson junction can be described, in the most general approach, by an Hamiltonian $\cal H$, which is a function of the phase difference $\varphi$ between the superconductive electrodes: 
\begin{equation}\label{H}
{\cal H}=-4E_{c} \frac{\partial^{2}}{\partial\varphi^{2}}-E_{J}\cos\varphi
\end{equation}


where $E_{c}$ and $E_{J}$ are the aforementioned charging and Josephson energies respectively.\cite{tinkham} $E_{c}$ is commonly much smaller than $E_{J}$, both in the HTS and in the LTS case, therefore 
the $E_{c}$ term in equation (1) is usually disregarded. In this condition, the dynamics of the junction phase can be modelled as the motion of a fictitious particle of mass \textit{m}=\textit{C}($\Phi_0$/2$\pi$)$^{2}$ in the "washboard" potential $U$($\varphi$)=$-E_{J}$[cos$\varphi$+(\textit{I}/\textit{I$_0$})$\varphi$], sketched in Figure 2.  This dynamics is well understood, both in the classical and in the quantum regime.\cite{bp, devoret_PRL85} For $\textit{I}<\textit{I$_0$}$ the potential $U$ has local minima where the phase particle is trapped and oscillates at the plasma frequency $\omega_0=\sqrt{2\pi I_0/C \Phi_0}$. An increase of $I$ has the effect of tilting the potential and decreasing the barrier between two neighbouring minima. Eventually, for $I=I_0$ the phase will escape from the well and a voltage will appear at the junction's edges. Decreasing the bias current, the potential tilt will be reduced and for $I=I_r$ the particle will be retrapped in a well, returning to the zero voltage state.
\\
In the case of underdamped junctions, with quality factor $Q_0=\omega_0RC>1$, we find $I_r<I_0$, therefore an hysteresis is present in the $I-V$ characteristic. In the case of overdamped junctions ($Q_0<1$), only one stationary state, the one at rest at a potential minimum with zero voltage across the junction, is stable for $\textit{I}<\textit{I$_0$}$ and the $I-V$ characteristics show no hysteresis.\cite{bp}
\\
This picture is strictly valid only at zero temperature. At finite temperature, thermal noise activates the phase over the energy barrier, favoring a slip from the potential well for $I=I_c<I_0$ (red line in Figure 2). In underdamped junctions, a single phase slip event is enough for the junction to switch to the running state. In overdamped junctions, on the other hand, after thermal slippage, the phase can be recaptured in the next well (blue line in Fig. 2). This prevents the  access to the running state and leads to the appearance of a non zero voltage, manifesting as a "rounding" in the $I-V$ curve at low currents. This regime is called phase diffusion\cite{bp, tinkham}.
\\
\subsection{Frequency dependent damping model}
A more complete description of the Josephson phase dynamics can be achieved by incorporating the effects of the circuit the junction is embedded into. The effects of the external environment are taken into account through an additional quality factor $Q_1$.\cite{kautz_martinis,LTScircuits}
In the case of HTS-based junctions, this external circuit is intrinsic and partly hidden, because it is embedded in the GB and, in the case of off-axis biepitaxial junctions, in the (103) oriented electrode.\cite{bauch_science, rotoli_PRB07} The study of its contributions, as encoded in the damping of HTS devices, therefore becomes more challenging. 
\\
The effects of the embedding circuit become particularly interesting when $Q_1<Q_0$. At the plasma frequency $\omega_0$ (typically in the GHz range), the smaller quality factor $Q_1$ dominates the behavior of the whole system. The voltage state involving steady motion of the phase is instead dominated by the higher quality factor $Q_0$. Therefore, the system will exhibit a frequency dependent damping, which explains the coexistence of hysteresis and phase diffusion,\cite{kautz_martinis} as seen in our devices (Figure 1b and 1d).
\\
When $E_c$ is comparable with $E_J$, the $E_c$ term in equation (1) cannot be disregarded. Its presence leads to phase delocalization effects. The value of the ratio $x=E_c/E_J$ is a measure of how strongly the charging energy acts in delocalizing the phase, being related to the width $\delta \varphi$  of the phase wave function $\psi (\varphi)$: $\delta \varphi = (x)^{1/4}$. For $x<<1$, $\psi (\varphi)$ is a narrowly peaked function, the phase is localized and can be treated as a semi-classical quantity. For values of $x$ greater than 1/4, on the other hand, the phase variable is sufficiently delocalized that quantum fluctuations cannot be neglected and quantum uncertainty, especially at low temperatures, has to be taken into account \cite{iansiti}. Phase delocalization leads to an increase in the probability for the phase to escape from the potential well, both in the thermal and in the quantum regime. Multiple escape and retrapping result in a finite resistance $R_0$ at low voltage; in the quantum regime, the value of $R_{0}$ saturates due to freezing out of the thermal fluctuations.
\\

\subsection{Numerical model}
In order to model frequency dependent damping in our devices, we use a two $Q$'s model, following the work of Kautz and Martinis\cite{kautz_martinis} (K-M model). The circuit considered is shown in the inset of Figure 2.\cite{nota_PRB} Conservation of current at nodes and the Josephson equations imply the following normalized Langevin equations for the phase $\varphi$ and the voltage $V_b$ at the external circuit capacitance $C_s$:

\begin{eqnarray}\label{phi}
\ddot{\varphi}=Q_{0}^{-2}\cdot
((V_b-\dot{\varphi})(Q_0/Q_1-1)-\dot{\varphi}-\sin \varphi+
\\
+\gamma_b+\gamma_{n1}+\gamma_{n2})\nonumber
\\
\label{vv}
\dot{V_b}=\rho Q_{0}^{-2}\left((\dot{\varphi}-V_b)+\gamma_{n2}/
(Q_0/Q_1-1)\right)
\end{eqnarray}

\noindent  in the equations above, time is normalized to $\hbar/2eI_0R_j=\omega_0^{-1}/Q_0$ and currents to the critical current $I_0$; $Q_0=R_j\sqrt{2eI_0C_j/\hbar }=\omega_0R_jC_j$ and $Q_1=(1/R_j+1/R_s)^{-1}\sqrt{2eI_0C_j/\hbar }=\omega_0R_s C_j$. The term $(V_b-\dot{\varphi})(Q_0/Q_1-1)$ represents the normalized current through external load $R_s$,
$\rho=R_jC_j/R_sC_s$ is the time constant ratio, $\gamma_b$ is the normalized bias current and $\gamma_{n1}$, $\gamma_{n2}$ are the noise currents, associated with
the intrinsic resistor $R_j$ and the external resistor $R_s$ respectively. These are modelled as Gaussian stochastic processes with zero mean and variance given by:
\begin{equation}\label{inin}
    \langle \gamma_{nk}(t),\gamma_{nk}(t') \rangle\equiv\sigma_k^2\delta(t-t')=\alpha_k\frac{ 2k_B T}{E_j}\delta(t-t')
\end{equation}
with $\alpha_1=1$ and $\alpha_2=Q_0/Q_1-1$. This simple model is able to reproduce the main features of experimental results\cite{kautz_martinis} without the use of other parameters. Simulations of the Langevin equations have been made generating Gaussian noise by cernlib RANLUX routine \cite{ranlux}. Other details of the numerical integration can be found in Ref.[\onlinecite{rotoli_PRB07}]. In order to capture the phase diffusion regime in $I-V$ characteristics an average procedure was performed over 2000 or 3000 single $I-V$s, depending on temperature. Each single $I-V$ was generated by averaging over $2000$ time units.  Typical runs for simulations of Eq.s (2) and (3) will last from 2$\cdot$10$^6$ to 4$\cdot$10$^6$ normalized time units, i.e., 10$^5$ to 2$\cdot$10$^5$ plasma periods.
\\
In the next section, we will compare our experimental data with the frequency dependent damping model. The low temperature measurements of device 6W are then discussed in the framework of the quantum phase diffusion regime.  

\begin{figure}[t]
\centering
        \includegraphics[width=8.0 cm, height=9.0 cm]{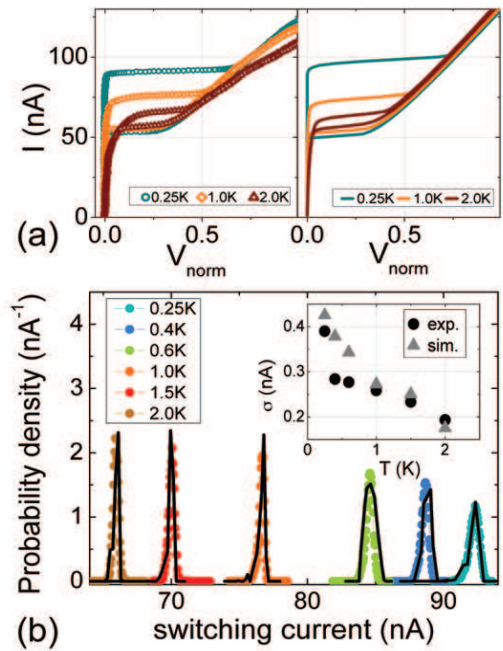} 
    \caption{(Color online) Transport properties of device 1W. In panel (a) the experimental $I-V$ characteristics (left part, points) measured at T=0.25K, 1.0K and 2.0K are compared with Monte-Carlo simulations (right part, full lines) realized using the following parameters: $Q_1$= 0.6, Q$_{0}$=5 and $I_0$=130nA. Panel (b) shows the comparison between experimental (points) and simulated (black full lines) SCD histograms. The experimental SCD histograms were measured using a voltage criterion of 100$\mu$V.  The inset shows the behavior of the simulated (triangles) and experimental (dots) histograms width versus the temperature.}
\end{figure}
 
\section{Comparison between numerical and experimental results}
Panel (a) of Fig. 3 shows the comparison between the experimental data of device 1W measured at different temperatures (left side) and numerical curves calculated using the two $Q$'s model (right side).
Significant changes in the shape of the experimental $I-V$ curves take place when cooling down from 2.0K, where the $I-V$'s exhibit a small hysteresis of 15$\%$ and a pronounced rounding of the low voltage branch, to 0.25K, where the hysteresis reaches 40$\%$ and a sharp switch from the superconducting to the resistive branch is observed. The simulations in the right hand side of Figure 3 reproduce this behavior well: the evolution of the critical current, the amplitude of the hysteresis and the coexistence of hysteresis and phase diffusion 'rounding'. The parameters used for the simulations are: $Q_{1}$=0.6$\pm$0.1, $Q_{0}$=5$\pm$0.5, $I_0$=130 nA. These are consistent with a capacitance per unit area of 1.5$\times$10$^{-6}$ Fcm$^{-2}$ (as observed in wider junctions\cite{stornaiuolo_SUST}), $\rho=0.1$ and an effective resistance of 500 Ohm. The experimental $I_c$ measured at 0.25K is only 70$\%$ of the $I_0$ value used for the simulations. This difference arises since the small $E_J$ means that, at 0.25K, $k_BT/E_J\sim1/10$ and so thermal noise currents (whose amplitude is proportional to  $\sqrt{k_BT/E_J}\sim0.31$) have a significant effect.
\\
For this device, we have measured the SCD histograms at various temperatures, reported in panel (b) of Figure 3. The standard deviation $\sigma$ of the experimental SCD histograms decreases as the temperature increases (dots in the inset of Figure 3b), as expected in the phase diffusion regime. The ratio between $\sigma$ and the mean switching current is in the range 10$^{-3}$, in agreement with that found in the literature.\cite{vion}
In Figure 3(b) we also show the fits to the SCD histograms (full lines). These were realized using the following parameters: $Q_{1}$=0.56, $Q_{0}$=2, $I_0$=130 nA. The switching behavior of a JJ is a high frequency phenomenon. Indeed, the study of the switching behavior of JJs in the moderately damped regime\cite{phase_diffusion, Longobardi_PRL12} is usually performed using a single-$Q$ model to fit the experimental SCD histograms. Such procedure works well when the condition $E_J>>k_BT$ is satisfied and the quality factor is larger than one. In our case, $Q_1$=0.56, therefore, in order to preserve the underdamped dynamics of the phase after the escape process, a second quality factor $Q_0$ with a slightly increased value with respect to $Q_1$, had to be included in the model.
\\
We point out that experimental reports showing the occurrence of phase diffusion effects both in the $I-V$ curves and in the SCD histograms are extremely rare. This combined analysis has previously been carried out, to our knowledge, only in Ref.[\onlinecite{vion}] where, contrary to what happens in our work, the main contribution to the damping of the devices comes for the  external impedance, and the junction intrinsic resistance plays no significant role. In our case, the reduced value of $E_J$ makes phase diffusion effects become evident not only in the behavior of the SCD histograms, but also in the shape of the $I-V$ characteristics, thereby offering two independent routes for the study of phase diffusion. An estimation of the high frequency dissipation $Q_1$ for our device, for instance, is both an output of the K-M model and a necessity for numerically reproducing the experimental SCD histograms. 
Finally, we point out that, in previous experiments on off-axis biepitaxial junctions realized on LSAT substrates, the $Q$ factor obtained via the simulation of SCD histograms was 1.3$\pm$0.05.\cite{Longobardi_PRL12} This value is consistent with $Q_1$=0.6$\pm$0.1 found in the present work, taking into account that here $I_c$ is one order of magnitude smaller and that high frequency dissipation is larger for devices with reduced $I_c$.\cite{confronto_bauch} 
 
\begin{figure}[t]
\centering
        \includegraphics[width=8.0 cm, height=5.0 cm]{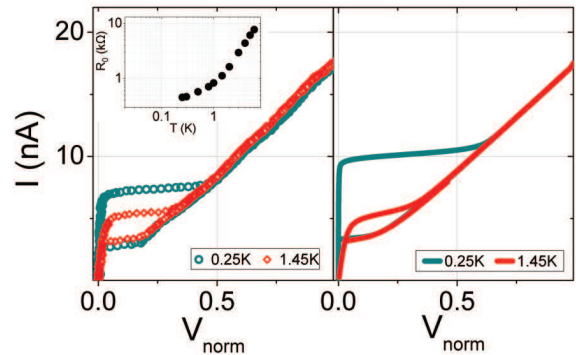} 
    \caption{(Color online) $IV$ characteristics of device 6W (points, left panel) measured at T=0.25K, and 1.45K compared with Monte-Carlo simulations (full lines, right panel) made using $Q_1$= 0.6 and $Q_0$=12 and $I_0$=35nA. In the inset the temperature dependence of the measured low voltage resistance $R_{0}$ is shown.}
\end{figure}
 
The analysis of the behavior of junction 1W reveals that the phase dynamics of YBCO submicron JJs characterized by low values of $E_J$ is compatible with that expected, in the K-M approach, in the phase diffusion regime. A further reduction of $E_J$, making it comparable to $E_c$, induces a different behavior, as we will demonstrate for device 6W.
\\
In Fig. 4 we compare the experimental $I-V$ curves of device 6W (left panel) with simulations (right panel). In this case, it was impossible to find a single set of parameters which could reasonably reproduce the $I-V$ curves in the complete range of temperatures. Agreement with the main features of the experimental data is obtained at high temperature (T=1.45K) by using the following parameters: $Q_1$= 0.6$\pm$0.1, $Q_0$=12$\pm$0.5 and $I_0$=35nA.\cite{comparisonQ1} Remarkable deviations appear as the temperature is reduced to 0.25K. We attribute such deviations to a transition from a classical regime, in which thermal fluctuations dominate, to a quantum regime in which phase delocalization plays a key role in the dynamics. Indeed, for this device $x=E_c/E_J$ is $0.65$ ($E_c\approx47\mu$eV, see Table I), leading to a region where phase delocalization effects are expected to be relevant, and promoting quantum phase diffusion.\cite{iansiti, QPD} 

The reduced value of $I_c$ of device 6W (a factor of 10 lower compared to device 1W) is consistent with this estimation of the fundamental energies. As discussed in the previous section, $x$ is related to the width of the phase function $\delta \varphi $ and therefore to the delocalization of the phase. For $x\approx 0.65$, $\delta \varphi $ is $\approx 0.9$. Although the phase $\varphi$ is still confined in one well of the washboard potential, the barrier height of such a well, which depends on both $E_J$ and $E_c$, is reduced, influencing the critical current. For $x>1/4$, the critical current $I_c$ is indeed scaled by $E_B$/$E_J$ where $E_B$ is the binding energy:\cite{iansiti}  
\begin{equation}
E_B \thickapprox E_J2x[(1+1/8x^2)^{1/2}-1]
\label{eq:5}
\end{equation}
leading to a temperature-independent $I_c=2eE_B/\hbar$ which is less than the value $I_0=2eE_J/\hbar$ which would be observed in the absence of quantum fluctuations. Using the values of $E_J$ and $x$ to calculate $E_B$, we obtain $I_c$=6.5nA, in good agreement with the experimental value measured at 0.25K (see Fig. 4). 
\\
As mentioned in Section III, the temperature dependence of the finite resistance at low voltages, $R_0$, is another indicator of the quantum phase diffusion state. Device 6W clearly shows such resistance, also at 0.25K, as marked by the black line in Figure 1d. Iansiti et al.\cite{iansiti} report that the value and the behavior of $R_{0}$ depends on the ratio $x$. The $R_{0}$ values shown in the inset of Fig. 4 are consistent with those found in Ref.\onlinecite{iansiti} resulting from numerical simulations using $x$=0.65.
Moreover, $R_0$ is proportional to the tunnelling rate\cite{iansiti} $R_0\approx \frac{h}{2eI} \Gamma$ and $\Gamma$ can be calculated by using the Caldeira-Leggett approximation in presence of dissipation\cite{Caldeira_Legget}. Using this formula with an upper bound value of $R_0\approx$500 Ohm, a damping $Q$ of about 1 is obtained. This value is consistent with the high frequency $Q_1$ factor inferred for this device (at high temperatures) using $I-V$ simulations. More importantly, the $R_0$ of device 6W decreases with decreasing temperature and levels off around 0.3K, as shown in the inset of Figure 4. The saturation of $R_0$ marks the entrance into the quantum regime.\cite{nota_SCD_6W} 
\\
\begin{table}[t]
	\centering
	\begin{tabular*}{0.45\textwidth}{@{\extracolsep{\fill}} c  c  c  c  c  c  c  c }
	
	\hline
	device	&	$E_J$	&	$E_c$	&	x	&	$Q_0$	&	$Q_1$	&	$I_0$ \\ \hline
	1W	&	270$\mu$eV	&	45$\mu$eV	&	0.16	&	5$\pm$0.5	&	0.6$\pm$0.1	&	130nA \\ \hline
	6W	&	70$\mu$eV	&	47$\mu$eV	&	0.65	&	12$\pm$0.5	&	0.6$\pm$0.1	&	35nA \\ \hline
	\hline
	\end{tabular*}
	\caption{Parameters of devices 1W and 6W. All the parameters refer to T=0.25K, except for $Q_0$, $Q_1$ and $I_0$ of device 6W, which refer to T=1.45K.}
\end{table}
From the estimated value of
the plasma frequency $\omega_0$ $\approx$ 40GHz, we calculate a crossover temperature $T_{cr}=\hbar \omega_0 / 2 \pi k_B$ between the classical
and the quantum regime, of 120mK.\cite{Tcross} Such equation for the crossover temperature has been estimated in the regime $E_J>>E_c$. In our case, since $E_J \approx E_c$, the binding energy is modified, the phase delocalization is larger and therefore the probability for quantum tunnelling of the phase is increased. As a result, the crossover temperature between thermal and quantum activation is pushed up. Indeed our experimental data show that quantum tunneling of the phase influences the phase dynamics already at 0.3K. 

We point out that junction 1W has
similar values of $\omega_0$ and $T_{cr}$ (75GHz and 155mK respectively)
but the condition $E_c<<E_J$ (see the values listed in Table I) results
in negligible delocalization effects, and the dynamics of
the junction is classical down to 0.25K, as shown by the
good agreement between the experimental data and the simulations (Fig. 3).

\section{Conclusions}
We have engineered YBCO grain boundary biepitaxial junctions in the submicron scale, down to 600nm, and with reduced Josephson energy $E_J$. This regime is quite rare to achieve for HTS JJs and has been, up to now, scarcely explored. The junctions behavior can be simulated using a frequency dependent damping model. The quality factors obtained by the fits indicate a moderately damped regime\cite{stornaiuolo_SUST, stornaiuolo_JAP, bauch_PRL}. Classical phase diffusion, in a frequency dependent approach, describes quite well the behavior of the devices, as far as $E_c<<E_J$. When $E_J\approx E_c$ delocalization starts to play an important role in the phase dynamics, the temperature at which quantum effects start to influence the phase dynamics is increased and a transition to a quantum phase diffusion regime occurs at T$\approx$0.3K.
\\
This work is of relevance both to define phase dynamics in HTS JJs in extreme limits and for the experimental search for quantum phase diffusion. More systematic studies will be required to obtain additional hints on the effects of microscopic factors, in particular the relation between a d-wave order parameter symmetry and dissipation.

\begin{acknowledgements}
We acknowledge the support of
MIUR-Italy through PRIN project 2009 “Nanowire high critical temperature superconductor field-effect devices” and COST project "Nanoscale Superconductivity: Novel Functionalities through Optimized Confinement of Condensate and Fields"
\end{acknowledgements}


\begin{thebibliography}{99}
\bibitem {oxygenLSAT} The submicron junctions studied in this work have lower $J_c$ when compared with the ones studied in Ref.\onlinecite{stornaiuolo_JAP}. This difference is likely due to the different fabrication process.
\bibitem{nota_PRB} The K-M model was just set to study phase-diffusion as a natural simplest extension of single Q model. This cannot be achieved with a "shell" circuit approach, as for example the one reported in Ref.\onlinecite{rotoli_PRB07}, which was implemented to model high frequency response of YBCO junctions realized on STO substrates.
\bibitem{confronto_bauch} In Ref.\onlinecite{bauch_PRL} the $Q$ of off-axis biepitaxial junctions was found to be 3.7$\pm$1.1. This higher value is likely due to the fact that the junctions were realized on a STO substrate.
\bibitem{QPD} Quantum phase diffusion arises when $E_c$ is comparable with $E_J$. This regime can be obtained by reducing either the $I_c$ value or the capacitance of the junction. In our case, the devices 1W and 6W are defined on the same chip and the capacitance can reasonably assumed to be constant.
\bibitem{nota_SCD_6W} A more quantitative analysis on the quantum phase diffusion regime for this device could be performed using SCD histograms. However, due to the extremely low values of the critical currents, such measurements are extremely difficult to perform. As a matter of fact we are not aware of SCD measurements in this range of critical currents, even for LTS JJs.
\end{thebibliography}
\end{document}